# Anthropogenic actinides in seawater and biota from the west coast of Sweden

M. López-Lora[1,2,*], E. Chamizo[1], M. Eriksson[3]

[1] Centro Nacional de Aceleradores (CNA), Universidad de Sevilla, Junta de Andalucía, Consejo Superior de Investigaciones Científicas, Thomas Alva Edison 7, 41092 Sevilla, Spain

[2] Departamento de Física Atómica Molecular y Nuclear, Universidad de Sevilla, Avenida de la Reina Mercedes s/n, 41012 Sevilla, Spain

[3] Department of Health, Medicine and Caring Sciences (HMV), Linköping University, 58183 Linköping, Sweden

* Corresponding author. mlopezlora@us.es

## Abstract

The assessment of the origin of the anthropogenic contamination in marine regions impacted by other sources than global fallout is a challenge. This is the case of the west coast of Sweden, influenced by the liquid effluents released by the European Nuclear Reprocessing Plants through North Sea currents and by Baltic Sea local and regional sources, among others. This work focused on the study of anthropogenic actinides ($^{236}$U, $^{237}$Np and $^{239,240}$Pu) in seawater and biota from a region close to Gothenburg where radioactive wastes with an unknown composition were dumped in 1964. To this aim, a radiochemical procedure for the sequential extraction of U, Np and Pu from biota samples and the subsequent analysis of $^{236}$U, $^{237}$Np, $^{239}$Pu and $^{240}$Pu by Accelerator Mass Spectrometry was developed. The method was validated through the study of two reference materials provided by the International Atomic Energy Agency (IAEA): IAEA-446 (Baltic Sea seaweed) and IAEA-437 (Mediterranean Sea mussels). The $^{233}$U/$^{236}$U atom ratio was also studied in the seawater samples. The obtained results indicate that the North Sea currents and global fallout are the major sources for $^{236}$U, $^{237}$Np and $^{239,240}$Pu to the studied area, with any clear evidence of other local sources. Complementary, information on the Concentrations Factors (CF) in biota was obtained, for which the available information is very scarce. For seaweed, CF values of $(4.1 \pm 1.1)\cdot10^3$, $88 \pm 51$ and $70 \pm 20$ have been obtained for Pu, Np and U, respectively. Lower CF values of $(3.6 \pm 1.2)\cdot10^2$, $37 \pm 15$ and $15.1 \pm 5.2$ for Pu, Np and U, respectively, have been obtained for mussels.

## Highlights

- Study of seawater and biota at Gothenburg, in the west coast of Sweden
- Assessment of contamination inputs by the study of multiple actinide radionuclides
- Development of a sequential method for the study of actinides in biota
- $^{236}$U, $^{237}$Np, $^{239,240}$Pu and $^{233}$U(seawater) analyses by Accelerator Mass Spectrometry
- Estimation of the Concentration Factors for U, Np and Pu in a real marine area

## 1. Introduction

The Baltic Sea is a complex marine region when it comes to assessing the presence of anthropogenic radioactivity sources. The reason is that the Baltic Sea has been historically exposed to diverse sources: the global fallout (GF) mainly from the atmospheric testing of thermonuclear weapons in the 1950s and 1960s, the atmospheric debris from the Chernobyl accident in 1986, and the liquid effluents from European Nuclear Reprocessing Plants (NRPs) entering Kattegat through North Sea currents, among others. The most important source-term to the Baltic Sea is GF, with estimated inventories of about 15 TBq for the combined activity of the two mayor plutonium isotopes, i.e. $^{239}$Pu ($T_{1/2}$ = 2.41·10$^4$ y) and $^{240}$Pu ($T_{1/2}$ = 6.56·10$^3$ y), 2 GBq (i.e. about 0.9 kg) for $^{236}$U ($T_{1/2}$ =2.34·10$^7$ y), and 50 GBq (i.e. about 2 kg) for $^{237}$Np ($T_{1/2}$ = 2.14·10$^6$ y) (Chamizo et al., 2015; Holm, 1995; Kelley et al., 1999). As for the Chernobyl influence on the Baltic Sea basin, about 1.5 TBq of $^{239+240}$Pu has been estimated (Holm, 1995), but with a very uneven distribution (Olszewski et al., 2018). The North Sea waters entering Kattegat, fingerprinted with NRPs liquid effluents, feature an increase content of the conservative radionuclides like $^{236}$U and $^{237}$Np (Castrillejo et al., 2020; OSPAR Commission London (U. K), n.d.; Scottish Environment Protection Agency (SEPA), n.d.), but the actual associated inventories remain unknown. Overall, $^{236}$U budgets up to 174 g and 323 g have been estimated in seawater and sediments, respectively, from nuclear reactor sources in Baltic Sea (Lin et al., 2022). Finally, in addition to those inputs, being surrounded by multiple nuclear developed countries, the Baltic Sea is exposed to other potential local and/or regional sources, as e.g. the controlled or accidental discharges from nuclear facilities (e.g., historical plutonium releases from the Swedish nuclear facility of Studsvik have been recently studied (López-Lora et al., 2023)). On top of that, there is a wide number of radioactive dumping sites reported all around the Baltic seabed, being the most of them not well-documented.

The study of source terms in complex marine regions like the Baltic Sea require the combined study of multiple actinide radionuclides, with source-dependent isotopic composition and different geochemical behaviors. Due to the particle reactive nature of Pu, the Pu isotopic composition has been the most widely used tool to this end specially in sedimentary reservoirs, being the $^{240}$Pu/$^{239}$Pu atom ratio in close connection with the contamination source (i.e. $^{240}$Pu/$^{239}$Pu atom ratios range from 0.01 from the first nuclear bomb test (Parekh et al., 2006) to above 0.57 for the Chernobyl accident (Bisinger et al., 2010), with a GF average signal of 0.18 ± 0.14 (Kelley et al., 1999)). In the last few decades, $^{236}$U and $^{237}$Np, both exhibiting a mostly conservative behavior in seawater, have entered the scene as tools to evaluate water masses. The $^{237}$Np/$^{236}$U atom ratio can be used to distinguish water masses fingerprinted by NRPs from those tagged by the GF signal (López-Lora et al., 2021). $^{237}$Np/$^{236}$U ratios of 1.1 – 1.7 and 0.02 – 0.3 are characteristic of liquid wastes released during the last two decades from Sellafield and La Hague NRPs, respectively, and a GF $^{237}$Np/$^{236}$U ratio of 1.77 ± 0.20 is estimated for seawater (López-Lora et al., 2020). Very recently, several studies of the most minor uranium isotope $^{233}$U ($T_{1/2}$ = 1.59·10$^5$ y) have set the basis to the use of $^{233}$U/$^{236}$U for disentangling anthropogenic uranium sources. Unlike $^{236}$U, $^{233}$U is not produce in significant amounts in conventional nuclear reactors ($^{233}$U/$^{236}$U at the level of 10$^{-6}$ is expected in burned nuclear fuel) and its main origin was a series of nuclear detonations with a specific design (i.e. using $^{233}$U as the fissile material or thermonuclear devices containing a tamper of $^{235}$U) (Hain et al., 2020). The most reliable estimation for the $^{233}$U/$^{236}$U GF ratio, based on a peat core from Austria, is $(1.40 \pm 0.15) \cdot 10^{-2}$ (Hain et al., 2020). To date, Accelerator Mass Spectrometry (AMS) offers the most competitive limits of detection for the analysis of $^{233}$U, $^{236}$U, $^{237}$Np and Pu isotopes in environmental samples (Kutschera, 2016).

To have an overview of the impact of a contamination source in a particular marine environment, the key is to study different compartments like seawater, sediments and biota. Marine sediments have been widely studied, being historical reservoirs of sedimented material that can be used to understand the fate of radionuclides over the time (Kuwabara et al., 1996; Lin et al., 2021; López-Lora et al., 2023). With the aim of assessing the transport of water masses, many studies have dealt with the analysis of conservative radionuclides like $^{236}$U in seawater (Castrillejo et al., 2017; Chamizo et al., 2017; Villa-Alfageme et al., 2019; Wefing et al., 2022). However, the situation has been different regarding marine biota. The reported data about the presence of anthropogenic actinides in samples such as seaweed, mussels and oysters are quite limited. Most of the available works have focused on the study of the $^{239+240}$Pu activities (Cross and Day, 1981; CROWLEY et al., 1990; Fisher et al., 1999; Holm, 1995; Howard et al., 2013; Ikäheimonen et al., 1997; Lindahl et al., 2005; Ryan et al., 1999), and very little information have been published regarding the presence of $^{236}$U and $^{237}$Np in biota (Chaplin et al., 2022; Germain et al., 1987; Lindahl et al., 2005; Pentreath and Harvey, 1981).

Actinides accumulate within the marine biota in concentrations several orders of magnitude higher than in seawater, presenting a potential radiological and biotoxicological risk for a marine ecosystem impacted by a local contamination source. Bivalve mollusks are filter-feeders, retaining particulates down to approximately 1-2 µm, and they live for a few decades. This, together with their wide distribution and abundance in all seasons in both fresh water and marine environments, has made them a good biological indicator of radionuclides adsorbed onto fine particulate matter (CROWLEY et al., 1990). Marine macroalgae do not only retrieve nutrients from the water but also efficiently absorb metals, including actinide radionuclides (Chaplin et al., 2022). The accumulation of actinides in the marine biota is described by the Concentration Factors (CF), i.e. the concentration of a specific radionuclide in the biological tissue (per unit mass of wet weight) divided by its concentration per unit mass in seawater. A range of recommended CF values have been reported by the International Atomic Energy Agency (IAEA) for U, Np and Pu (IAEA, 2004). However, a large dispersion of values for the same element and species have been found in multiples studies, being those IAEA CF reference values conditioned by physicochemical variations in the marine environment. Most of the current CF studies for U, Np and Pu are based on laboratory experiments which could not be representative of a real-case scenario. Therefore, more work is required for a reliable determination and understanding of the CFs for those elements (Chaplin et al., 2022).

Gothenburg, located at Kattegat, in the west coast of Sweden, is a coastal area characterized by the presence of an archipelago with numerous small islands surrounded by shallow waters with complex current systems mainly governed by wind. Gothenburg port is located at the mouth of the river Göta älv and, therefore, this area is influenced by the freshwater inflow from the river. Until it was banned in the 1970s, this shallow area was used as spoil ground, so it was considered particularly suitable for the dumping of waste and contaminated materials. A dumping of low-level nuclear waste in 1964 in this area has been documented. Those wastes were coming from contaminated demolition materials because of a uranium irradiation incident at Chalmers University (Gothenburg,

Sweden) in 1963. However, the information about this incident and the sea dumping is very limited and consists of unpublished reports from the SSM (Swedish acronym of Swedish Radiation Safety Authority). The exact coordinates of the site have not been reported and there is no information on activity content other than that the contaminated material contains radium and uranium (but plutonium and fission byproducts are also likely to be found in the waste). The status of this dumped radioactive material is not known, and the existence of a possible leakage has not been studied up to now.

In this work, we aimed for the first time to the analysis of $^{236}$U, $^{237}$Np and $^{239,240}$Pu in seawater and biota (seaweed, blue mussels and oysters) samples from Gothenburg to find evidence for leakages from the dumped material in 1964. To this end, a new sample preparation method for biota samples has been developed and tested using two IAEA reference materials: IAEA-446 (Baltic Sea seaweed) and IAEA-437 (Mediterranean Sea mussels). To complement this study, $^{233}$U/$^{236}$U ratios have also been determined from seawater samples. Thus, the final goals of this work are i) to identify the contamination sources and to clarify if there is an impact from the nuclear dumped material at Gothenburg, ii) to provide a reliable method for the determination of $^{236}$U, $^{237}$Np and $^{239,240}$Pu by AMS in biota and iii) to perform a direct estimation of CF values for biota in a real marine environment.

## 2. Materials and Methods
### 2.1. Samples and studied area

Figure 1 shows the selected sampling sites for seawater and biota around Gothenburg in this work. Although the exact location of the dumping site in 1964 remains unknown, some reported information pointed towards the surroundings of the small Danska Liljan island (Fig.1, stations 2 and D). Therefore, the adopted sampling strategy included stations around the Danska Liljan island, and a further location considered as a background station (Fig. 1, stations 1 and A). Two sampling campaigns were carried out in October 2020 and August 2021. During the last campaign a more limited number of samples were collected due to logistical problems.

Surface seawater samples of 10 L were collected from 4 stations (Fig. 1) and filtered onboard (0.45 µm pore size). Samples were then acidified to pH < 2 with hydrochloric acid and set aside for one week for homogenization purposes. 40 mL aliquots from the 10 L samples were then extracted to quantify $^{238}$U by Ionization Coupled Plasma Mass Spectrometry (ICP-MS).

Live seaweed samples (*Fucus serratus*) of about 3-5 kg (wet weight) were collected from 4 stations (Fig. 1). The life span expected for those seaweed species is 3 – 5 years. They were dried at 90°C and homogenized. Sub samples of about 5 g (dry weight) were taken for actinides analysis.

Live blue mussels (*Mytilus edulis*) and live oysters (*Magallana gigas*) were collected from 3 stations (stations 1, 2 and 4, Fig. 1). Those specimens can reach up to about 20 – 30 years of age but their average survival span is about 2 – 3 years. They were dissected to extract the soft tissues and combined according to the collection site for each specie. Samples were then freeze dried and homogenized. Sub samples of about 5 g (dry weight) were taken for actinides measurements.

Together with those samples, the applied methods were validated through the study of two reference materials provided by the IAEA: IAEA-437 (Mediterranean mussel (Pham et al., 2010)) and IAEA-446 (Seaweed from the Baltic Sea (Pham et al., 2014)).

**2.2. Radiochemical methods**

2.2.1. Materials

New glassware and reagents of the highest purity were used to prevent background issues. Ion chromatography separations were performed by using 2 mL cartridges (TEVA®, UTEVA® and Pre-filter® resins, provided by Eichrom Technologies Inc.) in a vacuum box. For the $^{236}$U, $^{237}$Np and $^{239,240}$Pu AMS measurements, the following spikes were used: $^{233}$U ($^{236}$U) was provided by IAEA (unknown supplier), and $^{242}$Pu ($^{237}$Np and $^{239,240}$Pu) was provided by NPL (National Physical Laboratory, United Kingdom). The $^{242}$Pu and $^{233}$U spike solutions have been previously measured for the absence of $^{239,240}$Pu, $^{237}$Np and $^{236}$U traces (i.e. below the $10^{-5}$ atomic abundance in each solution). For the validation of the proposed procedure, a $^{237}$Np standard solution provided by the IAEA was used. A $^{244}$Pu standard solution provided by Join Research Center (European Commission) was used to prepare the secondary solution used in this work to quantify the chemical recoveries by AMS.

2.2.2. Seawater

Seawater samples were chemically processed for the AMS analysis of $^{233}$U, $^{236}$U, $^{237}$Np, $^{239}$Pu and $^{240}$Pu following the sequential method reported in (López-Lora et al., 2019) and schematized in Figure 2. Actinides were first pre-concentrated with $Fe(OH)_2$ and then Pu+Np and U were separated using TEVA® and UTEVA® resins in tandem, respectively. Only $^{242}$Pu was added as an initial spike. Note that this method relies on the use of $^{242}$Pu as a non-isotopic tracer for $^{237}$Np. The $^{238}$U naturally present in the samples, measured by ICP-MS from independent aliquots, was used to estimate the $^{236}$U and $^{233}$U concentrations from the $^{236}$U/$^{238}$U and $^{233}$U/$^{238}$U final AMS ratios.

2.2.3. Biota samples: seaweed, mussels and oysters

Subsamples of about 5 g (dry weight) from biota samples (i.e., seaweed, mussels and oysters) were taken for the analysis of $^{236}$U, $^{237}$Np, $^{239}$Pu and $^{240}$Pu by AMS and processed following the proposed method (Fig. 2). They were calcinated in a muffle furnace at 550°C for 10 hours and about 1-2 pg of $^{242}$Pu and $^{233}$U were added as spikes to quantify the final concentrations of $^{239}$Pu, $^{240}$Pu and $^{237}$Np ($^{242}$Pu spike) and $^{236}$U ($^{233}$U spike). Following the guidelines of the procedure for seawater samples in (López-Lora et al., 2019), $^{242}$Pu was used as a yield tracer for $^{237}$Np. Samples were leached on a hot plate at 120°C, first with aqua regia for about 8 hours and then with 8M $HNO_3$ and $H_2O_2$ for about 6 hours. Then, they were treated with concentrated HCl, dissolved in about 3mL of HCl and diluted with MQ water up to about 0.5 – 1 L. 2 g of potassium metabisulphite ($K_2S_2O_5$) and 200 mg of Fe(II) as iron sulfate ($FeSO_4$) were added to the samples. The $Fe(OH)_2$ precipitate was obtained by increasing the pH to 9 with concentrated ammonia ($NH_4OH$). U and Pu fractions were purified and sequentially separated by using TEVA® and UTEVA® resins in tandem following the same procedure than for seawater samples (López-Lora et al., 2019). For actinides AMS analysis, the final purified samples were adapted to the optimal matrix for the AMS analysis as explained in (López-Lora et al., 2019).

## 2.3. AMS analyses of $^{233}$U, $^{236}$U, $^{237}$Np, $^{239}$Pu and $^{240}$Pu

Actinide analysis of $^{233}$U (only in seawater), $^{236}$U, $^{237}$Np, $^{239}$Pu and $^{240}$Pu (seawater and biota) were performed by using the 1 MV AMS system at the Centro Nacional de Aceleradores (CNA, Seville, Spain) (Chamizo et al., 2021; Chamizo and López-Lora, 2019; López-Lora and Chamizo, 2019). Helium was used as stripper gas and the tandem accelerator was set at about 650 kV. In-house secondary standard materials were used to normalize the measured atomic ratios (i.e. $^{236}$U/$^{233}$U, $^{236}$U/$^{238}$U, $^{239,240}$Pu/$^{242}$Pu, $^{237}$Np/$^{242}$Pu) (Calvo et al., 2015; López-Lora and Chamizo, 2019). The detection limit of the technique is at the $10^5$-$10^6$ atom level and an abundance sensitivity of about $1\cdot10^{-10}$ is reached for the $^{236}$U/$^{238}$U ratios. All the uncertainties presented in this article are expressed with k = 1 (68%) confidence level.

## 2.4. ICP-MS analysis of $^{238}$U and $^{235}$U in seawater samples.

In collaboration with SSM, $^{238}$U and $^{235}$U isotopes were measured by using an Agilent 8900 ICP-MS/MS instruments (Agilent Technologies) at the SSM in Stockholm and at Linköping University facilities (Sweden). Thus, $^{238}$U and $^{235}$U concentrations and $^{235}$U/$^{238}$U isotope ratios were quantified from seawater samples following the technique reported in (Lindahl et al., 2021).

## 3. Validation of the method for seaweed and mussel samples
### 3.1. Pretreatment

For biota samples, after the calcination at 550ºC, a leaching in a hot plate was performed to extract the radionuclides (Fig. 2). To test the efficiency of this pretreatment step as well as its reproducibility, several replicates were processed to test different leaching options (Supplemental material, Figure S1). Finally, a doble leaching on a hot plate at 120ºC was stablished: a first leaching with aqua regia (6 hours) followed by a 8M HNO$_3$ leaching adding H$_2$O$_2$ (6 hours).

### 3.2. Reliability of the method for $^{237}$Np

The reliability of using a non-isotopic tracer (i.e. $^{242}$Pu) for $^{237}$Np analysis was specifically tested for biota samples. To this end, aliquots of seaweed and blue mussel samples were spiked with a known amount of $^{237}$Np and $^{242}$Pu, i.e. "control samples". The results from the measured concentrations of $^{237}$Np together with the expected values (added $^{237}$Np concentrations) are shown in Figure 3 for set of control samples (5 control samples of seaweed and 2 of mussels). There was a good agreement in all the samples considering uncertainties, resulting in an average ratio of the measured over the added $^{237}$Np concentrations of 1.04 ± 0.08 for seaweed and 0.93 ± 0.07 for mussels.

### 3.3. Radiochemical yields

Radiochemical yields for the proposed procedure were obtained by AMS following the method reported in (Chamizo et al., 2024). 5% aliquots of the final Pu+Np and U solutions were spiked with a secondary tracer ($^{244}$Pu) to quantify the final amount of the initial spike (i.e., $^{242}$Pu and $^{233}$U, respectively) in that fraction and, by extrapolation, in the whole AMS sample. The obtained recoveries were reasonable high in all the samples (seaweed and blue mussels), being above 72% and above 65% for Pu and U, respectively.

### 3.4. Validation of the method using IAEA biota reference materials

To validate the proposed method for biota samples, two IAEA reference materials were analyzed (Figure 2): IAEA-446 (Baltic Sea Seaweed (Pham et al., 2014)) and IAEA-437 (Mussels from the Mediterranean Sea (Pham et al., 2010)). The results from those samples are summarized in Table 1.

In the case of IAEA-446, the obtained $^{239}$Pu atom concentration agrees with the reported values, but the $^{240}$Pu atom concentration and $^{239+240}$Pu activity concentration are slightly below the published ranges. Still, there is a reasonably agreement considering the wide range of values. The obtained $^{240}$Pu/$^{239}$Pu atom ratio is 0.166 ± 0.012, which is below the 0.214 – 0.226 interval provided in (Pham et al., 2014). Nevertheless, looking at the row data provided by different laboratories (ICP-MS and AMS techniques) detailed in (*Worldwide Laboratory Comparison on the Determination of Radionuclides in IAEA-446 Baltic Sea Seaweed (Fucus vesiculosus)*, 2013), the reported $^{240}$Pu/$^{239}$Pu atom ratios are quite variable, ranging from 0.17 to 0.33. This could point out to inhomogeneities in this sample or possible inconsistencies in the reported $^{240}$Pu levels. On the other hand, the $^{236}$U concentration of (37.7 ± 0.9)·10$^9$ atoms/kg (i.e. 35.4 ± 0.8 nBq/g) obtained in this work is 2-3 orders of magnitude below the values provided by two ICP-MS facilities (i.e. 3.4 ± 0.2 µBq/g and 59 ± 6 µBq/g) in (*Worldwide Laboratory Comparison on the Determination of Radionuclides in IAEA-446 Baltic Sea Seaweed (Fucus vesiculosus)*, 2013). The $^{236}$U concentration obtained from IAEA-446 in this work is similar to other results from Baltic seaweed samples also in this work (Table 2). An overestimation of $^{236}$U in those previous ICP-MS analysis might explain those discrepancies. No previous information has been reported for $^{237}$Np from the IAEA-446.

Regarding IAEA-437, the $^{239+240}$Pu activity obtained in this work is in reasonably agreement but slightly below the reported range in (Pham et al., 2010) from alpha spectrometry analysis. The independent analysis of $^{239}$Pu and $^{240}$Pu have been previously provided by solely one laboratory (ICP-MS technique (Pham and Sanchez-Cabeza, 2007)), reporting a range of 0.183 – 0.205 for the $^{240}$Pu/$^{239}$Pu atom ratio, which is close to the obtained value in this work (i.e. 0.173 ± 0.017). Our results approach the concentrations ranges provided for $^{239}$Pu and $^{240}$Pu by ICP-MS. To our best knowledge, no information about neither $^{236}$U nor $^{237}$Np has been previously reported for the reference material IAEA-437.

In conclusion, the method proposed in this work for biota samples (Figure 2) has been proved to be a reliable method for the sequential determination of $^{236}$U, $^{237}$Np, $^{239}$Pu and $^{240}$Pu by AMS for seaweed and mussel samples. Because of the very little material that we were able to sample in the case of oysters, no specific tests were performed in those samples. The final method validated for seaweed and mussels is expected to be suitable for oysters, but it has not been specifically tested.

### 4. Study of Gothenburg's marine environment
### 4.1. Seawater

AMS results for the seawater samples collected in the two sampling campaigns (October 2020 and August 2021, Figure 1) are detailed in Table S1 and summarized in Table 2 and figure 4. ICP-MS results for the natural uranium isotopes $^{238}$U and $^{235}$U are included in Table S1. $^{238}$U concentrations ranging from 1.18 µg/kg to 2.41 µg/kg, with an average value of 1.84 ± 0.17 µg/kg, where obtained. All the seawater samples showed a $^{235}U/^{238}U$ ratio in agreement with the natural expected ratio (i.e. 0.00724). Regarding activity concentrations, similar values have been found in all the studied stations during the two sampling campaigns, ranging from 1.2 to 2.2 µBq/kg for $^{239+240}Pu$ (i.e. (1.8 – 1.6)·10$^6$ atoms/kg for $^{239}Pu$ and (0.13 – 0.26)·10$^6$ atoms/kg for $^{240}Pu$), from 36·10$^6$ to 71·10$^6$ atoms/kg for $^{237}Np$ (i.e. 0.37 – 0.73 µBq/kg) and from 27·10$^6$ to 53·10$^6$ atoms/kg for $^{236}U$ (i.e. 25 – 50 nBq/kg). Moreover, $^{236}U/^{238}U$ atom ratios are in the range of (8.7 – 9.3)·10$^{-9}$.

Fluctuations and variations in 2020 with respect to 2021 are most probably due to variations in the water current in this shallow area. Comparable $^{239+240}Pu$ activities have been reported at Kattegat from seawater samples collected in 2008 (i.e. 1.2 – 3.0 µBq/kg) ("MARIS - Marine Radioactivity Information System," n.d.). Previous $^{237}Np$ results in the area were in the range of (110 – 116)·10$^6$ atoms/kg and (45 – 100)·10$^6$ atoms/kg from samples collected in 1999-2001 and 2008, respectively (Lindahl et al., 2005; "MARIS - Marine Radioactivity Information System," n.d.). $^{237}Np$ results obtained in this work confirm the decreasing trend for this radionuclide in seawater in the area. This is also in agreement with the decline $^{237}Np$ emissions from Sellafield NRP (OSPAR Commission London (U. K), n.d.), which is, at first, the main contribution to $^{237}Np$ signals together with the GF (i.e. the presence of $^{237}Np$ in liquid discharges from La Hague NRP is several orders of magnitude below Sellafield releases). Regarding $^{236}U$, the obtained $^{236}U$ concentrations and $^{236}U/^{238}U$ atom ratios fit within the range of values reported in previous works from surface seawater samples collected at Kattegat in 2013 – 2018, i.e. $^{236}U$ concentrations of (33 – 105)·10$^6$ atoms/kg and $^{236}U/^{238}U$ ratios of (2.4 – 20)·10$^{-9}$ (Lin et al., 2022; Qiao et al., 2021).

All the obtained $^{240}Pu/^{239}Pu$ atom ratios agree considering the uncertainties. They fit within the expected values in the area, mostly governed by GF (i.e. 0.180 ± 0.014 (Kelley et al., 1999)) and the current inflow of waters from the North Sea, labelled by Sellafield and La Hague releases (i.e. 0.16 – 0.24 (Lindahl et al., 2011)). $^{237}Np/^{236}U$ ratio in the area is expected to be influenced by the same main inputs, being 1.77 ± 0.20 the reported GF ratio (López-Lora et al., 2020) and 1.1 – 1.7 the estimate range of values for the combined inputs of Sellafield and Lague releases during the last two decades (López-Lora et al., 2021). The obtained $^{237}Np/^{236}U$ ratios in this work, ranging from 1.0 to 1.5, are within those expected values. Finally, $^{233}U/^{236}U$ ratios agree for all the studied samples considering the uncertainties, resulting in an average value of about 0.35·10$^{-2}$. This result fits within the range of values reported in previous works for seawater samples collected at Kattegat in 2013 – 2018, i.e. (0.2 – 0.5)·10$^{-2}$ (Lin et al., 2022; Qiao et al., 2021).

Therefore, seawater results from this work are within the expected levels from this area for all the studied radionuclides and agree with the main expected inputs (i.e. GF and North Sea input labeled by releases from NRPs of Sellafield and La Hague). The sampling area is also potentially exposed to possible leakage from radioactive waste materials dumped in 1964. Although the exact location of those dumped radioactive wastes is not known, it is most likely to be place around station D and 2 (Figure 1). The results from this work do not reflect any significant indication of a possible leakage from those radioactive dumped materials.

### 4.2. Biota

AMS results from biota samples (seaweed, blue mussels and oysters) collected in the two sampling campaigns performed in 2020 and 2021 are shown in Figure 5 and Table S1, and the average values summarized in Table 2. Seaweed samples were collected from 4 stations, but blue mussels and oysters were only found in 3 of them. Moreover, very little material was found in the case of oysters and only some information about $^{236}$U was obtained. $^{236}$U/$^{238}$U ratios from biota were at the level of $1\cdot10^{-8}$, approaching the obtained ones from seawater samples. This could point out to a high efficiency of the leaching for natural U isotopes. However, since no total digestion have been performed during the sample preparation for biota, those values are not considered for the discussion.

For seaweed samples, the obtained concentrations (dry weight) have been in the range of 30 – 42 mBq/kg for $^{239+240}$Pu, (8 – 41)·10$^9$ atoms/kg for $^{237}$Np (i.e. 82 – 420 µBq/kg) and (15 – 20)·10$^9$ atoms/kg for $^{236}$U (i.e. 14 – 19 µBq/kg). Considering the results from the water samples in the area obtained in this work (Figure 5), the higher uptake of particle-reactive plutonium from seaweed compared to the more conservative uranium and neptunium is evidenced. Moreover, $^{239+240}$Pu and $^{237}$Np are within the range of values reported from seaweed samples collected at Kattegat in a previous work (Lindahl et al., 2005). To our best knowledge, no previous results from $^{236}$U in seaweed samples have been reported neither in the Baltic nor in the North Sea regions.

For blue mussel samples, very low concentrations (dry weight), of about 2.8 – 5.1 mBq/kg for $^{239+240}$Pu, (10 – 15)·10$^9$ atoms/kg for $^{237}$Np (i.e. 34 – 61 µBq/kg) and (3.3 – 5.9)·10$^9$ atoms/kg for $^{236}$U (i.e. 3 – 6 µBq/kg ) were found. Similar $^{236}$U values were also obtained from oysters, i.e. (4.3 – 5.3)·10$^9$ atoms/kg or 3 – 6 µBq/kg, presenting a good agreement between their average values (see Table 2). $^{239+240}$Pu results from this work are 1 – 2 orders of magnitude lower than previous values reported from mussels collected in areas highly impacted by Sellafield or La Hague NRPs (CROWLEY et al., 1990; Germain et al., 1987; Ryan et al., 1999). Similarly, a higher value of (1.9 ± 0.5)·10$^{12}$ atoms/kg for $^{237}$Np was reported from mussels collected near La Hague (Germain et al., 1987).

Regarding the atom ratios results, very homogenous values have been obtained for the $^{240}$Pu/$^{239}$Pu, with average values of 0.1906 ± 0.0010 and 0.209 ± 0.014 for seaweed and mussels, respectively. Interestingly, those values are higher than the average value from seawater (i.e. 0.165 ± 0.006). Seaweed and mussels are expected to be reflecting the average signal of the last 2 – 5 years. This could indicate a slight tendency towards a decrease in the $^{240}$Pu/$^{239}$Pu ratio in the area. Anyway, those values are all compatible with the major expected inputs in the area and fluctuations are expected due to the combination of water currents. Some variations were found in the ratios from the conservative radionuclides. There is a slight decrease in the $^{237}$Np/$^{236}$U ratio from seaweed at station 2 that might be caused by the water mixing patterns in the area. On average, $^{237}$Np/$^{236}$U ratios from seaweed are compatible with the seawater results (i.e. average values of 1.21 ± 0.07 and 1.1 ± 0.3 have been obtained for seawater and seaweed, respectively). Mussels have showed a higher $^{237}$Np/$^{236}$U ratio of 2.48 ± 0.17, which might reflect a higher bioavailability of $^{237}$Np in mussels compared to $^{236}$U as it has been pointed out in a previous study in oyster shells (Huang et al., 2024). However, since only two results of the $^{237}$Np/$^{236}$U ratio have been obtained from mussels, more data will be needed to confirm this hypothesis. In contrast, $^{236}$U/$^{239}$Pu ratios are higher in seawater than in biota, because of the higher plutonium uptake in biota. Thus, average $^{236}$U/$^{239}$Pu values of 35 ± 5, 0.74 ± 0.03 and 1.8 ± 0.3 have been obtained for seawater, seaweed and mussels, respectively. Some differences have also been observed in mussels and seaweed samples, being the average $^{236}$U/$^{239}$Pu in mussels above the value from seaweed.

Finally, assuming no significant variations in the area, the average values in seawater and biota samples (Table 2) have been used to estimate the corresponding CFs values (Table 3). For those calculations, the concentrations per unit of wet weight mass of the biota samples have been considered following the guidelines for CF calculations reported by IAEA (IAEA, 2004). The CF values for seaweed obtained in this work for Pu, Np and U agree with the reported recommended values by IAEA. CF calculations for U in mussels and oysters in this work approach the IAEA CF value for U in mollusks. In contrast, the CF estimations from mussels in this work for Pu and Np are about one order of magnitude below the recommended IAEA CF for mollusks as well as the estimated one for Pu in (Howard et al., 2013). Nevertheless, it is important to notice that those recommended values are estimations based on a few studies solely and a very large dispersion of values have been reported in several studies (e.g. Pu CF in seaweed from previous works, Table 2). Therefore, the observed variations in the data for mussels might be caused by the mollusk species or the specific conditions of the marine environment.

## 5. Conclusions

This work presents a study about the presence of $^{236}$U, $^{237}$Np and Pu isotopes (i.e. $^{239}$Pu and $^{240}$Pu) from seawater and biota samples collected at the marine environment of Gothenburg, at the west coast of Sweden. Additionally, complementing the study, the $^{233}$U/$^{236}$U ratios have been obtained from seawater. The results reveal the GF and North Sea water inflow (i.e. labeled by Sellafield and La Hague radioactive discharges) as the main sources in the area. Locally, the potential input from radioactive low-level waste materials dumped in the area in 1964 has been also evaluated. No significant evidence of leakage affecting this marine environment has been found. In this work, a new radiochemical method has been established for the sequential extraction of uranium, neptunium and plutonium from biota samples for AMS determinations of $^{236}$U, $^{237}$Np, $^{239}$Pu and $^{240}$Pu. The procedure is based on the use of a non-isotopic tracer (i.e. $^{242}$Pu) for $^{237}$Np and it has been tested in seaweed and mussel matrices. The results have proved the reliability of the method, and it has been successfully applied for the analysis of two IAEA reference materials. Although the method was not initially intended for $^{233}$U determinations in biota (because of the low levels expected), the $^{236}$U results obtained in this work point to a promising future application of the method for $^{233}$U studies in biota. This work provided a better understanding of the radioactive levels in biota for the understudied radionuclides $^{236}$U and $^{237}$Np and demonstrated the great potential of those long-lived actinides for marine studies and source terms determinations.

**Acknowledgments**

This work has been supported by the Swedish Radiation Safety Authority (SSM) through the project SSM2019-9660. This project has received funding from the European Union's Horizon Europe Research and Innovation program under Grant Agreement No 101057511 (EURO-LABS) and from Grant PID2022-140680NB-I00 funded by MCIN/AEI/10.13039/501100011033 and by the European Union. M. López-Lora acknowledges the funding received from European Union's Horizon Europe–Marie Skłodowska Curie HORIZON-MSCA-2022-PF-01 through the Postdoctoral Fellowship with Ref. 101106068 (CHRONUS). Authors would like to thank Bohusläns Musseum and Göteborg Energi for providing background information and let us study the sonar survey material for the identification of potential dumped objects in the area, and to Marie Karlsson, Per Törnquist and Jan Scholten for their assistant during the sampling campaigns.

**Tables**

*Table 1. Results from the reference materials IAEA-446 (Baltic Sea Seaweed (Pham et al., 2014)) and IAEA-437 (Mussels from the Mediterranean Sea (Pham et al., 2010)). The reported values in previous works are considered information values, only the $^{239+240}$Pu activity concentration from IAEA-446 is a certified value. \* Range for AMS and ICP-MS values reported in (Worldwide Laboratory Comparison on the Determination of Radionuclides in IAEA-446 Baltic Sea Seaweed (Fucus vesiculosus), 2013). \*\* ICP-MS values reported in (Pham and Sanchez-Cabeza, 2007).*

|  |  | $^{239}$Pu ($10^9$ at/kg) | $^{240}$Pu ($10^9$ at/kg) | $^{239+240}$Pu (mBq/kg) | $^{240}$Pu/$^{239}$Pu (atom ratio) | $^{236}$U ($10^9$ at/kg) | $^{237}$Np ($10^9$ at/kg) |
|---|---|---|---|---|---|---|---|
| IAEA-446 (Seaweed) | This work | 14.5 ± 0.5 | 2.40 ± 0.15 | 21.3 ± 0.5 | 0.166 ± 0.012 | 37.7 ± 0.9 | 13.3 ± 0.6 |
|  | *Reported values* | *13 – 16* | *2.7 – 4.2* | *22 – 26* | *0.214 – 0.226 (0.17 – 0.33)\** |  |  |
| IAEA-437 (Mussels) | This work | 4.13 ± 0.17 | 0.71 ± 0.06 | 6.2 ± 0.3 | 0.173 ± 0.017 | 44.0 ± 1.0 | 59.2 ± 1.3 |
|  | *Reported values* | *4.4 – 5.5 \*\** | *0.8 – 1.1 \*\** | *7.1 – 8.2 (6.4 – 9.0)\*\** | *0.183 – 0.205 \*\** |  |  |

Table 2. Average values for the all the samples collected in this work at the stations detailed in Figure 1 during the two sampling campaigns (2020 and 2021). Average concentrations for biota have been calculated considering both the initial dry and wet weight of the samples.

|  |  | $^{239+240}$Pu (mBq/kg) | $^{237}$Np ($10^9$ atoms/kg) | $^{236}$U ($10^9$ atoms/kg) | $^{240}$Pu/$^{239}$Pu | $^{237}$Np/$^{236}$U | $^{236}$U/$^{239}$Pu |
|---|---|---|---|---|---|---|---|
| **Seawater** |  | (1.79 ± 0.14)·10$^{-3}$ | 0.060 ± 0.005 | 0.047 ± 0.003 | 0.165 ± 0.006 | 1.21 ± 0.07 | 35 ± 5 |
| **Seaweed** | (dry w.) | 36.5 ± 2.7 | 18 ± 5 | 18.1 ± 1.0 | 0.1906 ± 0.0010 | 1.1 ± 0.3 | 0.74 ± 0.03 |
|  | (wet w.) | 7.3 ± 1.5 | 3.6 ± 1.3 | 3.6 ± 0.7 |  |  |  |
| **Mussels** | (dry w.) | 3.8 ± 0.5 | 13 ± 3 | 4.7 ± 0.5 | 0.209 ± 0.014 | 2.48 ± 0.17 | 1.8 ± 0.3 |
|  | (wet w.) | 0.60 ± 0.19 | 2.0 ± 0.5 | 0.75 ± 0.15 |  |  |  |
| **Oysters** | (dry w.) | - | - | 4.8 ± 0.3 | - | - | - |
|  | (wet w.) | - | - | 0.62 ± 0.08 |  |  |  |

*Table 3. Concentration factors (CF) for biota estimated in this work together with the IAEA recommended values (IAEA, 2004) and other reported values in previous works. CF estimations from this work has been calculated using the average values from all the studied samples (Table 2).*

|  |  | Concentration Factor (CF) | | |
|---|---|---|---|---|
|  |  | **Pu** | **Np** | **U** |
| Seaweed | This work | $(4.1 \pm 0.9) \cdot 10^3$ | $61 \pm 22$ | $76 \pm 16$ |
|  | *IAEA recommended values* | *$4 \cdot 10^3$* | *50* | *100* |
|  | Other works | $4.3 \cdot 10^3$ (Lindahl et al., 2005) | 35 (Lindahl et al., 2005) | 50 – 300 (Chaplin et al., 2022) |
|  |  | $(1.6 \pm 0.7) \cdot 10^4$ (Fisher et al., 1999) |  |  |
|  |  | $(1.5 – 3.5) \cdot 10^3$ (Chaplin et al., 2022) |  |  |
| Mussels | This work | $(3.4 \pm 0.8) \cdot 10^2$ | $34 \pm 10$ | $16 \pm 3$ |
| Oysters | This work | - | - | $13.1 \pm 1.9$ |
| Mollusks | *IAEA recommended values* | *$3 \cdot 10^3$* | *400* | *30* |
|  | Other works | $10^3$ (Howard et al., 2013) |  |  |

## Figures

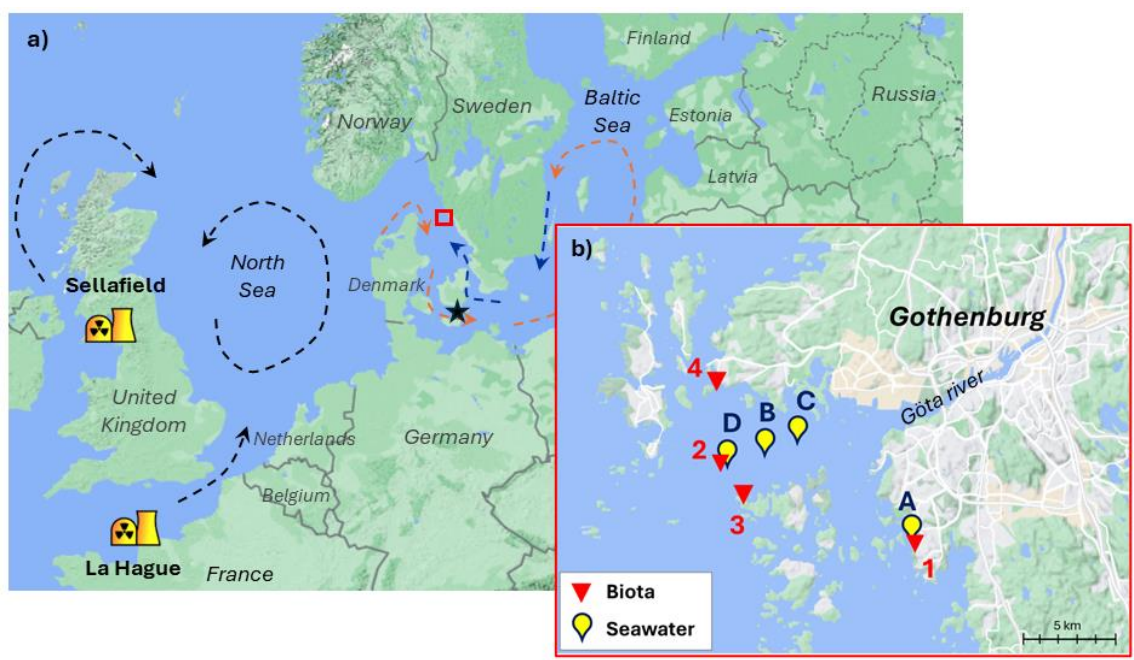

*Figure 1.- **Study region and sampling map.** Overview of the schematic circulation of water mass in the North Sea-Baltic Sea region (a). The black arrows sketch the water mass flow in the North Sea region. The orange arrows refer to the entrance of saline bottom waters in the Baltic Sea and the blue arrows the surface water outflow. The nuclear reprocessing plants of Sellafield (United Kingdom) and La Hague (France) are indicated in the map. The black star shows the sampling location for the reference seaweed sample IAEA-446. Expanded map of the studied region (b), in the marine area around the Göta river mouth, Gothenburg (Sweden). The sampling stations for seawater and biota (seaweed and/or mussels and oysters) are indicated.*

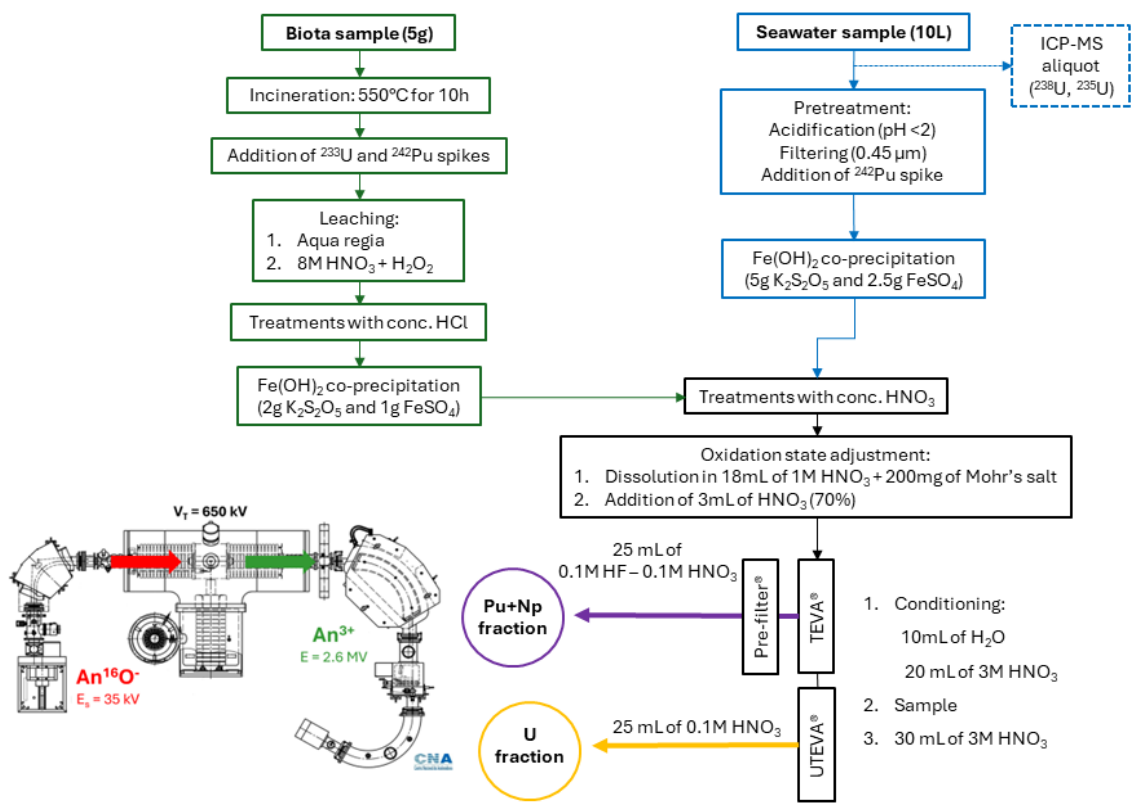

*Figure 2.- **Flow diagram of the radiochemical procedure for seawater and biota samples and the AMS analysis at the 1MV AMS system at the CNA**. Proposed work for biota samples (this work) based on the reported method for seawater samples in (López-Lora et al., 2019). AMS analyses were intended for $^{236}$U, $^{239}$Pu, $^{240}$Pu and $^{237}$Np from biota and seawater samples. In the case of seawater, $^{233}$U measurements were also performed, therefore, no U spike was added during the sample preparation. In this case, aliquots of the initial seawater samples were measured by ICP-MS to estimate the $^{236}$U and $^{233}$U concentrations from the naturally present $^{238}$U and $^{235}$U in the samples.*

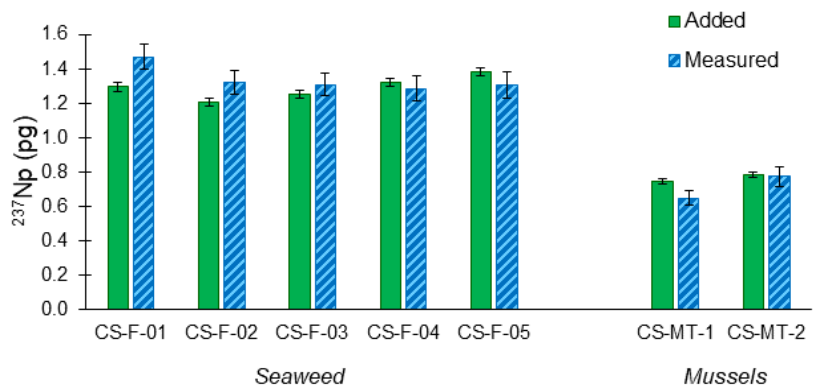

*Figure 3.- **Results from "control samples"**. $^{237}$Np concentration added and measured from "control samples", i.e. calcinated samples spiked with a known amount of $^{237}$Np to test the reliability of the method. Seaweed samples (CS-F) and mussel samples (CS-MT) were tested.*

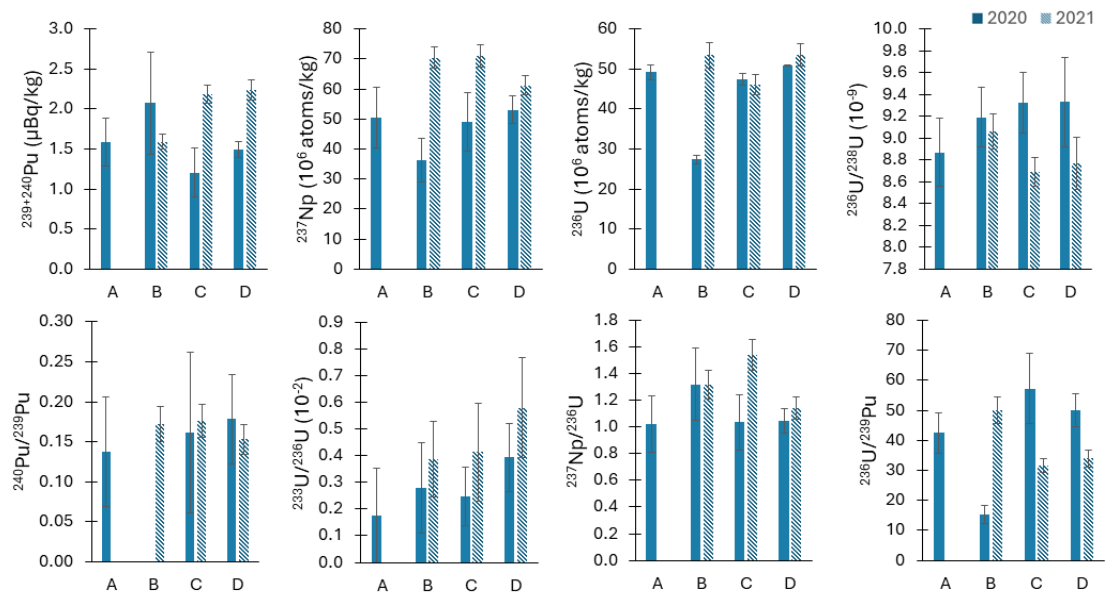

*Figure 4: AMS results from seawater.* Samples collected in 2020 and 2021. The location of the different stations (A, B, C and D) is shown in Figure 1. $^{236}U/^{238}U$, $^{240}Pu/^{239}Pu$, $^{233}U/^{236}U$, $^{237}Np/^{236}U$ and $^{236}U/^{239}Pu$ refer to atom ratios.

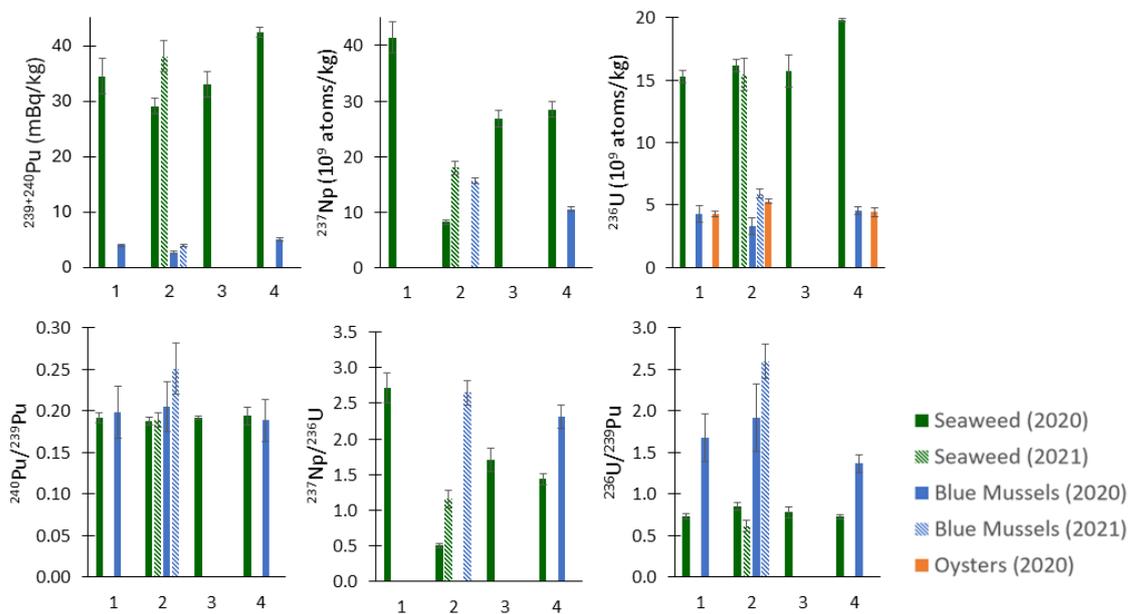

*Figure 5: AMS results from biota.* Samples of seaweed, blue mussels and oysters collected in 2020 and 2021. The locations for the different stations (1, 2, 3 and 4) are shown in Figure 1. Concentrations are calculated from dry weight. $^{240}Pu/^{239}Pu$, $^{237}Np/^{236}U$ and $^{236}U/^{239}Pu$ refer to atom ratios.

# Supplementary material

**Table S1 –**

Results obtained in this work by AMS from seawater, seaweed and mussels samples collected at the stations detailed in Figure 1. For seawater samples, ICP-MS results for the determination of the natural uranium isotopes are also included. Available at https://github.com/AMS-CNA/Actinides-biota-seawater-Sweden/

**Pretreatment for biota samples**

For biota samples, after the calcination at 550ºC, a leaching in a hot plate was performed to extract the radionuclides. In order to test the efficiency of this pretreatment step as well as its reproducibility, several replicates were processed to test different leaching options. Aliquots of about 5 g (dry weight) from the same seaweed/mussel sample were used to this end. The results are shown in Figure S1, where "MT" refers to mussel aliquots and "FT" to seaweed. The tested leaching methods were the following:

- Leaching 1: Samples were leached on a hot plate with 8M $HNO_3$ + $H_2O_2$ for 6 hours at 120°C (MT-A, MT-B, FT-A, FT-B in Fig. 3)
- Leaching 2: Samples were leached on a hot plate at 120°C with aqua regia for 6 hours (MT-C, MT-D, FT-C, FT-D in Fig.3).
- Leaching 3: Samples were leached on a hot plate at 120°C with aqua regia for 6 hours and then with 8M $HNO_3$ + $H_2O_2$ for 6 hours (MT-E, MT-F, FT-E, FT-F, FT-G).

The results from these tests did not show any significant variation in the measured $^{239+240}$Pu, $^{236}$U and $^{237}$Np concentrations. Therefore, we assume that the efficiency of the three methods is good enough to extract those anthropogenic radionuclides. For the sake of making a conservative choice, leaching 3 was chosen for the final method (Figure 2).

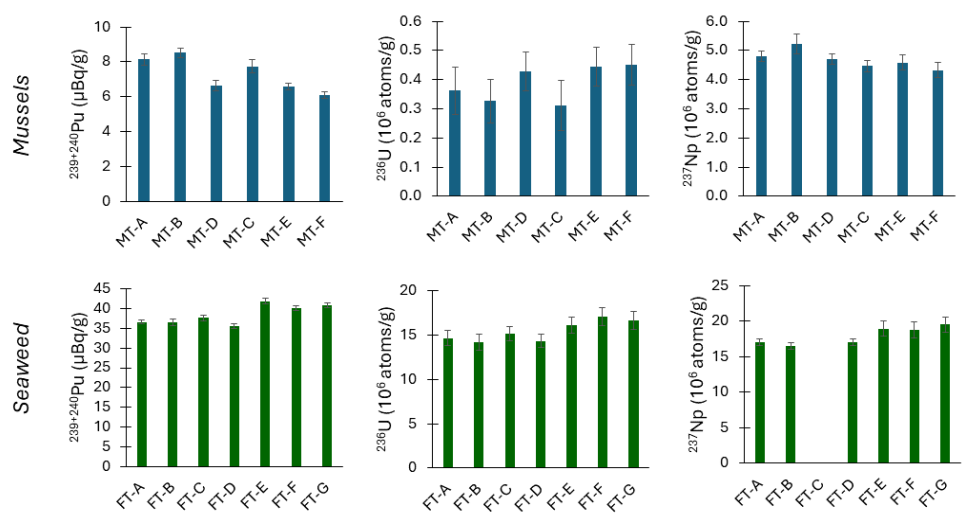

**Figure S1: Results from the leching test for $^{239+240}$Pu, $^{236}$U and $^{237}$Np.** Replicates from the same mussel (above) and seaweed (below) sample materials of about 5 g processed following different leching methods: (1) leaching with aqua regia (aliquots A, B), (2) leaching with 8M $HNO_3$ + $H_2O_2$ (aliquots C, D), and (3) leaching with aqua regia and 8M $HNO_3$ + $H_2O_2$ (aliquots E, F, G). $^{237}$Np concentration from sample FT-C is not considered because of a possible contamination issue.

19